\documentclass[aps,prx,amsmath]{revtex4}

\usepackage{times}
\usepackage{amsmath,amssymb}
\usepackage{graphicx}
\usepackage{color}
\usepackage{xspace}
\usepackage{mathtools}

\newcommand{\bra}[1]{\langle #1|}
\newcommand{\ket}[1]{|#1\rangle}
\newcommand{\braket}[2]{\langle #1 | #2 \rangle}

\newcommand{\smfrac}[2]{\mbox{$\frac{#1}{#2}$}}

\newcommand{\e}{\mathrm{e}}
\renewcommand{\i}{\mathrm{i}}

\newcommand{\pder}[2]{\frac{\partial #1}{\partial #2}}
\newcommand{\der}[2]{\frac{d #1}{d #2}}
\newcommand{\dder}[2]{\frac{d^2 #1}{d #2^2}}
\newcommand{\ppder}[2]{\frac{\partial^2 #1}{\partial #2^2}}

\newcommand{\RR}{\mathbb{R}}

\newcommand{\QAO}{\textsc{qao}\xspace}

\newcommand{\QMC}{\textsc{qmc}\xspace}

\newcommand{\opA}{\hat{A}}
\newcommand{\opB}{\hat{B}}
\newcommand{\opP}{\hat{P}}
\newcommand{\opM}{\hat{M}}
\newcommand{\opI}{\hat{I}}
\newcommand{\opQ}{\hat{q}}
\newcommand{\opx}{\hat{x}}

\newcommand{\gmin}{g_\mathrm{min}}

\renewcommand{\epsilon}{\varepsilon}

\preprint{Preprint}

\begin{document}

\title{Spectral Gap Analysis for Efficient Tunneling in Quantum Adiabatic Optimization}

\author{Lucas~T.~Brady}
\affiliation{{Department of Physics, University of California, Santa Barbara, CA 93106-5110, USA}}
\author{Wim~van~Dam}
\affiliation{{Department of Computer Science, Department of Physics, University of California, Santa Barbara, CA 93106-5110, USA}}

\date{\today}
\begin{abstract} 
We investigate the efficiency of Quantum Adiabatic Optimization when overcoming potential barriers to get from a local to a global minimum. Specifically we look at $n$ qubit systems with symmetric cost functions 
$f:\{0,1\}^n\rightarrow \RR$
where the ground state must tunnel through a potential barrier of width $n^\alpha$ and height $n^\beta$.  By the quantum adiabatic theorem the time delay sufficient to ensure tunneling grows quadratically with the inverse spectral gap during this tunneling process. We analyze barrier sizes with $1/2\leq \alpha+\beta$ and $\alpha<1/2$ and show that the minimum gap scales polynomially as $n^{1/2-\alpha-\beta}$ when $2\alpha+\beta \leq 1$ and  exponentially as $n^{-\beta/2}\exp(-C n^{(2\alpha+\beta-1)/2})$ when $1<2\alpha+\beta$.  Our proof uses elementary techniques and confirms and extends an unpublished folklore result by Goldstone, which used large spin and instanton methods.  Parts of our result also refine recent results by Kong and Crosson and Jiang et al.\ about the exponential gap scaling.  
\end{abstract}

\maketitle

\section{Introduction}
\label{sec:intro}

Quantum Annealing seeks to solve optimization problems by taking a state of a quantum system and evolving its Hamiltonian to get a desired result.  Quantum Adiabatic Optimization (\QAO) \cite{Farhi2000} is a form of quantum annealing that seeks to keep a system in the ground state while adiabatically evolving the Hamiltonian. A lot of recent work has gone into analyzing \QAO in its own right \cite{Farhi2002, Reichardt, Farhi2008} and comparing \QAO with classical algorithms such as Simulated Quantum Annealing via path-integral Quantum Monte Carlo \cite{Heim,Boxio2,Martonak,Hastings,Muthukrishnan,Brady,Crosson,Isakov} to see how much speed-up \QAO can give if any.

It has been conjectured that a large part of \QAO's power comes from the ability of quantum systems to tunnel through potential barriers. In this article, we focus on an $n$-qubit Hamiltonian, but by making it symmetric in the qubits, we can effectively reduce our problem to a one-dimensional tunneling problem.  This setup of one-dimensional tunneling in $n$ symmetric qubits has been studied before by Farhi, Goldstone, and Gutmann \cite{Farhi2002} who considered tunneling through a constant width spike of height $n$ and who showed for this setting a gap scaling of $\gmin \propto n^{-1/2}$. Reichardt \cite{Reichardt} showed that \QAO can tunnel in constant time ($\gmin\propto 1$) provided that the area (width $\times$ height) of the barrier is bounded by $O(\sqrt{n})$.  More recently, Crosson and Deng \cite{Crosson} examined thin barriers of varying height, and Kong and Crosson \cite{Kong} found that sufficiently large barriers lead to exponential run-times. Jiang et al.~\cite{Jiang} showed that Quantum Monte Carlo (\QMC) can reproduce the exponential run-time behavior of thermally assisted quantum tunnelling through such large barriers. Independently the current authors have numerically found \cite{Brady} that the transitions between constant, polynomial, and exponential run-time scaling for \QMC simulations coincide with the same transitions for \QAO. 

In this article, we consider barriers with width proportional to $n^\alpha$ and height proportional to $n^\beta$ and mainly focus on barriers with $1/2 \leq \alpha+\beta$, which is above Reichardt's \cite{Reichardt} constant scaling region, and $2\alpha+\beta<1$ which is below Kong ond Crosson's \cite{Kong} exponential scaling region. We show that barriers in this intermediate size regime lead to polynomial scaling of the minimum spectral gap with 
$\gmin \propto n^{1/2-\alpha-\beta}$. Through the quantum adiabatic theorem this scaling implies that a polynomial running time is sufficient for the \QAO algorithm to tunnel through such barriers. Additionally, our method also confirms Kong and Crosson's exponential scaling and provides an exact form for the polynomial prefactor on the exponential.

In section \ref{sec:ham}, we present our problem and discuss the Hamiltonian governing the interactions of our $n$-qubit system. Section \ref{sec:ana_gap} presents details of previous work on this problem and highlight both the polynomial scaling region between $1/2\leq \alpha+\beta$ and $2\alpha+\beta<1$ where few solid results have been published and the unexplored region for $\alpha>\beta$.

Our problem lends itself to a large spin analysis using either spin coherent states \cite{Auerbach} or the Villain transformation \cite{Villain}. Section \ref{sec:large_spin} briefly touches on spin coherent states, which have been used to analyze this problem before \cite{Kong}, and presents an in depth analysis using the Villain transformation, resulting in a semi-classical Hamiltonian that describes our dynamics for large $n$.

Focusing on just the critical region of the problem where the spectral gap is
smallest, section \ref{sec:quad_app} derives a model that approximates the
semi-classical Hamiltonian in the large $n$ limit. We provide several
arguments for why this model accurately represents the asymptotic behavior of
our original problem. Finally, in section \ref{sec:asymp}, we use this
model to derive an exact asymptotic expression for the scaling behavior of the
spectral gap.

\section{Quantum Adiabatic Optimization of Symmetric Functions}
\label{sec:ham}

Our main goal is to explore quantum tunneling through a barrier in a
symmetric cost function $f:\{0,1\}^n\rightarrow \RR$ defined on the
$n$-dimensional hypercube $\{0,1\}^n$. Our specific cost function is 
\begin{equation}
      f(x) = |x|+b(|x|),
\end{equation} 
where $|x|$ is the Hamming Weight of the length $n$ bit string $x$. The barrier
function, $b:\{0,\dots,n\}\rightarrow \RR$, is some function that is localized
around $|x|=n/4$ and has width proportional to $n^{\alpha}$ and height
proportional to $n^{\beta}$. We describe these barriers using the notation
$n^{\alpha}\times n^{\beta}$.

To create an algorithm to minimize the cost function, we first encode it into a 
quantum Hamiltonian on $n$ qubits:
\begin{equation}
H_1 = \sum_{x\in\{0,1\}^n}f(x)\ket{x}\bra{x}.
\end{equation}

\QAO starts the system in a different Hamiltonian with known and easily prepared
ground state; the typical starting Hamiltonian applies a magnetic field in the
$\hat{x}$ direction so that
\begin{align} 
      H_0 & = \sum_{i=1}^n (\mathsf{H}_0)_i \quad \text{with} \quad  
       \mathsf{H}_0  = \frac{1}{2}
      \left(\begin{array}{cc}1&-1\\-1&1\end{array}\right).
\end{align}

The ground state of $H_0$ is an equal superposition of all states $\ket{x}$,
which corresponds to a binomial distribution in Hamming weight. Then, \QAO finds
the ground state of $H_1$ by slowly evolving the system from $H_0$ into $H_1$
using
\begin{equation} \label{eq:Hs}
      H(s) = (1-s) H_0 +s H_1,
\end{equation}
and the quantum adiabatic theorem says that the system will stay in the ground
state if $0\leq s\leq 1$ varies slowly enough. In order to ensure adiabaticity
the evolution time, $T$, must scale depending on both the norm of $\der{H(s)}{s}$ and the inverse of the minimum
spectral gap between the two smallest eigenvalues $\lambda_1(s)$ and
$\lambda_0(s)$ of $H(s)$:
\begin{equation}
\gmin \coloneqq \min_{s\in[0,1]}\left(\lambda_1(s)-\lambda_0(s)\right).
\end{equation} 
Historically, sources \cite{Farhi2000} have claimed that the adiabatic theorem requires
\begin{equation}
T\gg \frac{\max_{s}||\der{H}{s}||}{\gmin^2}.
\end{equation}
Recent work \cite{Jansen} has shown that the adiabatic condition may be a more
complicated function of these parameters, but all of this recent work has the running time
scaling like an inverse polynomial in $\gmin$. Since the norm of the
Hamiltonian's derivative is usually independent of parameters such as our
$\alpha$ and $\beta$, typically the gap is taken as the important part of this
expression. Therefore, the key issue of this paper is the calculation of
$\gmin$.

The Hamiltonian, $H(s)$, on $n$ qubits can be simplified by considering just the
symmetric subspace. For each Hamming weight $0\leq h\leq n$, the Hamiltonian is
degenerate on the subspace of $\{\ket{x}: |x|=h\}$, so there will only be one
degenerate energy level for each Hamming Weight $h$. This symmetry can be
utilized to rewrite the full $2^n\times 2^n$ Hamiltonian as an
$(n+1)\times(n+1)$ symmetric Hamiltonian:
\begin{align}
      \label{eq:tridiag}
      H_\mathrm{sym}(s) =& \sum_{h=0}^n
            \left[\frac{(1-s)}{2}n+s(h+b(h))\right]\ket{h}\bra{h}\nonumber\\
            -&\frac{(1-s)}{2}\sum_{h=0}^{n-1}
                  \sqrt{(h+1)(n-h)}\ket{h}\bra{h+1}\nonumber\\ 
            -&\frac{(1-s)}{2}\sum_{h=0}^{n-1}
                  \sqrt{(h+1)(n-h)}\ket{h+1}\bra{h}
\end{align}

When $b(z)=0$, the ground state of the symmetric Hamiltonian is explicitly
\begin{align}
      \ket{GS_{b(z)=0}} =& \frac{1}{\left(2\delta(\delta+s)\right)^{n/2}}\nonumber\\
      \times&\sum_{h=0}^n \sqrt{n \choose h} (s+\delta)^{n-h}(1-s)^h\ket{h},
\end{align}
where $\delta\coloneqq \sqrt{1-2s+2s^2}$ is the unperturbed spectrl gap. This
distribution is a binomial for $s=0$, and the width remains proportional to
$\sqrt{n}$ for $0\leq s<1$. The maximum amplitude $\ket{h}$ state here
corresponds with the zero amplitude state in the first excited state and can be
thought of as the center of the distribution. The center coincides with
$h=\frac{n}{4}$ when $s = s^*\coloneqq\frac{1}{2}(\sqrt{3}-1)$.

In the large $n$ limit with a non-zero barrier, $b(z)$ becomes extremely narrow
relative to the dimension of the Hilbert space, so for most $s$ values, the
energy states are unaffected by the barrier. It is only when the energy states
get close to the barrier that the perturbation becomes important. Therefore, in
the large $n$ limit, the location of the minimum spectral gap becomes this
critical $s^*$.

\section{Previous Asymptotic Results}
\label{sec:ana_gap}
A folklore result by Goldstone \cite{Goldstone} says that for $\alpha<\beta$ and
$\alpha<1/2$ the minimum gap for tunneling through an $n^\alpha\times n^\beta$
barrier scales as a function of $n$ like
\begin{equation}
      \gmin \propto 
            \begin{cases}
                  1 & \text{if~}\alpha+\beta\leq\frac{1}{2}\\
                  n^{1/2-\alpha-\beta} & \text{if~}\alpha+\beta>\frac{1}{2}~\text{and}~2\alpha+\beta\leq 1\\
                  \text{poly}(n) \cdot \text{exp}\left(-C\,n^{(2\alpha+\beta-1)/2}\right) &\text{if~} 2\alpha+\beta>1.
            \end{cases}
      \label{eq:gold_pred}
\end{equation}

\begin{figure}
      \includegraphics[width=0.48\textwidth]{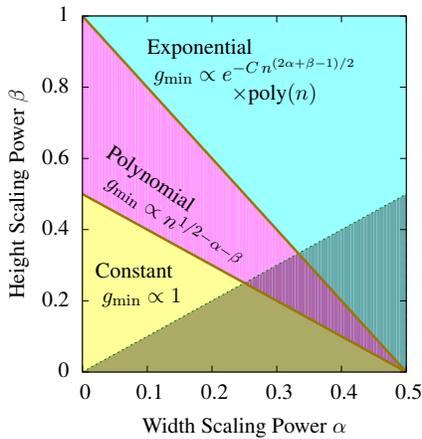}
      \caption{
            The spectral gap scaling of \QAO according to the
            original folklore result by Goldstone \cite{Goldstone}. This
            large $n$ behavior describes tunnelling through a barrier of size
            $n^\alpha\times n^\beta$ in the setting of $n$ symmetric qubits. The
            folklore result is restricted as it only works for $\alpha<\beta$ and
            $\alpha<1/2$, and it predicts constant, polynomial, or exponential
            scaling of the minimum gap $\gmin$ depending on the barrier size.
            The proof of this result has not been formally published.
      }
      \label{fig:regions_gold}
\end{figure}

\begin{figure}
      \includegraphics[width=0.48\textwidth]{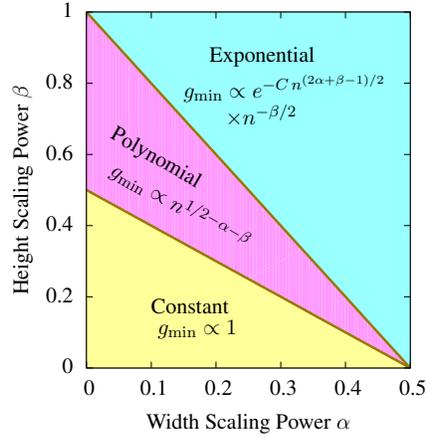}
      \caption{
           The spectral gap scaling during \QAO tunneling through a barrier of
           size $n^\alpha\times n^\beta$. Unlike Fig.~\ref{fig:regions_gold},
           this figure displays all current knowledge of each region, which
           includes the case $\alpha>\beta$. The yellow ``Constant'' region was
           proven by Reichardt \cite{Reichardt}, and the blue ``Exponential''
           region was shown in \cite{Kong, Jiang} up to the polynomial
           prefactor. The current article proves the polynomial scaling $\gmin
           \propto n^{1/2-\alpha-\beta}$ for the red region with
           $1/2<\alpha+\beta$ and $2\alpha+\beta<1$. Our article alse determines
           the polynomial prefactor for the blue exponential region described by
           $2\alpha+\beta>1$ with $\gmin\propto
           n^{-\beta/2}\exp(-Cn^{(2\alpha+\beta-1)/2})$.  
      }
      \label{fig:regions}
\end{figure}
While this result has never been published, its derivation is known to use
``large spin and instanton methods'' \cite{Goldstone}.
Fig.~\ref{fig:regions_gold} shows the scaling behavior according to
\cite{Goldstone}. Parts of Goldstone's result have been verified by several other
sources.

Reichardt \cite{Reichardt} rigorously proved the existence of the constant
region, and his results apply to the entire region where $\alpha+\beta<1/2$ not
just for $\alpha<\beta$. Recently, Kong and Crosson \cite{Kong} verified the
behavior of the gap in the exponential region for $2\alpha+\beta>1$ using the
instanton method \cite{Coleman}, and Jiang et al.~\cite{Jiang} have also found
the same exponential scaling behavior for the runtime of thermally assisted
quantum annealing on this barrier problem using a WKB approach. In a previous
article \cite{Brady} we numerically analyzed the transition between the
polynomial and exponential regions. Notably, no previously published work has
been able to verify the polynomial region, and while Kong and Crosson
\cite{Kong} proved the exponential region scaling, they restricted their proof
to $\alpha<\beta$ and did not derive the polynomial prefactor. The different
scaling regions in $\alpha$ and $\beta$ are shown in Fig.~\ref{fig:regions},
with references in the figure caption to which sources proved that region's
scaling behavior, including what is proven in this paper.

The goal of the current article is to explore both the polynomial region between
$\alpha+\beta>1/2$ and $2\alpha+\beta<1$ and in general the region where
$\alpha>\beta$. We develop elementary techniques to analyze the spectral
gap and verify the polynomial, $n^{1/2-\alpha-\beta}$, scaling behavior,
and we show that the results of Eq.~\ref{eq:gold_pred} are valid even when
$\alpha>\beta$.

\section{Large Spin  Approximation}
\label{sec:large_spin}

Our Hamiltonian readily lends itself to reinterpretation as the Hamiltonian for
a single particle with spin $J=n/2$. A common analytic technique for dealing
with a spin Hamiltonian is to use spin coherent states \cite{Auerbach} to create
a semi-classical continuous version of the Hamiltonian. Several groups
\cite{Farhi2002, Kong} have used spin coherent states to analyze qubit systems,
and Kong and Crosson \cite{Kong} have employed spin coherent states to analyze
the symmetric barrier problem for exponentially small gaps. We use a
similar technique employing a modified and formalized version of the Villain
transformation \cite{Villain}. The Villain transformation has been used for
similar problems \cite{Enz, Boulatov, Boxio}; in this article, we present a more
formal approach to this transformation in Appendix~\ref{app:villain}.

If we re-imagine our Hilbert space as representing a spin $J=n/2$ particle and
associate $\hat{J}_z$ eigenstates $\ket{m}$ with $\ket{h}$ states through
$\ket{m} = \ket{h-J}$, then our symmetric Hamiltonian can be rewritten in terms
of spin operators $\hat{J}_i$ as
\begin{equation}
      \label{eq:spin_ham1}
      \hat{H}(s) = -(1-s)\hat{J}_x+s\hat{J}_z+s\,b(\hat{J}_z+J) + J \Delta,
\end{equation}
where $\Delta$ represents some constant. Since we only care about energy
differences, this constant $\Delta$ can be arbitrary, and later, we use it
to ensure that the bottom of our potential energy well sits at zero energy.

Large spin techniques then pull a factor of $J= n/2 \eqqcolon 1/\epsilon$ out of
our Hamiltonian so that we are dealing with operators $\hat{\jmath}_i =
\epsilon\hat{J}_i$ that have eigenvalues that run from $-1$ to $+1$.
Specifically, we call the $\hat{\jmath}_z$ eigenvalue $-1\leq q\leq 1$, and
in the large $J$ (i.e.\ small $\epsilon$) limit, $q$ can be treated as a
continuous variable. We also introduce $r(q)\coloneqq \epsilon b(Jq+J)$
that is zero everywhere except in the vicinity of $q=-1/2$ where there is a bump
of width $\epsilon^{1-\alpha}$ and height $\epsilon^{1-\beta}$. Our Hamiltonian
can be rewritten as
\begin{equation}
      \label{eq:spin_ham2}
      \epsilon\hat{H}(s) = -(1-s)\hat{\jmath}_x+s\hat{\jmath}_z+sr(\hat{\jmath}_z) + \Delta.
\end{equation}
At this point, we can write an approximate Schr\"{o}dinger equation for this
Hamiltonian using the Villain transformation. In Appendix \ref{app:villain}, we
have taken the standard Villain transformation and made its logic more formal,
applying it specifically to Eq.~\ref{eq:spin_ham2}. In making
the logic more formal, we have held off taking the continuum limit of $q$ as
long as possible. The end result of the Villain transformation itself before
making any assumptions about the properties of our eigenstates gives a continuum
Schr\"{o}dinger equation:

\begin{align}
      \label{eq:schro_vill}
      \epsilon E \psi(q) &= \left(sq+sr(q)+\Delta-(1-s)\sqrt{1-q^2}\right.\nonumber\\
                            &\left.-(1-s)\frac{\epsilon^2}{2}\sqrt{1-q^2}\ppder{}{q}+ \mathcal{O}(\epsilon)\right)\psi(q).
\end{align}

The first line includes a potential energy, and the next one contains the
kinetic term for the problem. Note that the norm of the second derivative
operator, $\ppder{}{q}$, is proportional to $\epsilon^{-2}$ which is why this
term survives. At this point, the problem cannot be simplified without making
reference to the eigenstates we want to solve for. Notably, if we assume we are
at $s^*=\frac{1}{2}(\sqrt{3}-1)$ where the minimum of the potential energy is at
$q=-\frac{1}{2}$ in the $\epsilon\to0$ limit and make reasonable assumptions
about the nature of the ground state and first excited state, then
Eq.~\ref{eq:schro_vill} can be simplified even more. In
Appendix~\ref{app:villain}, we formalize these approximations, and in
Sec.~\ref{sec:quad_app} we analyze the resulting approximate differential
equation.

\section{Quadratic Potential Approximation}
\label{sec:quad_app}



In Appendix~\ref{app:villain}, we continue our approximation of
Eq.~\ref{eq:schro_vill} by noting that the low-lying energy states for
$s^*=\frac{1}{2}(\sqrt{3}-1)$ are centered in the extremely close vicinity of
$q=-1/2$. This allows us to focus on the variable $x\coloneqq q+\frac{1}{2}$ and
the region near $x=0$. For the low-lying energy states, such as the ground state
and first-excited state that we care about, the approximate differential
equation representing our problem in the small $\epsilon$ limit is

\begin{align}
      \label{eq:post_villain}
      \ppder{\psi}{x} &= \frac{1}{\epsilon^{2}}\left[\omega^2 x^2+{\smfrac{4}{3}}r(x-{\smfrac{1}{2}})-c \epsilon E\right]\psi(x),
\end{align}
where $c\coloneqq 8/(3(\sqrt{3}-1))$ and $\omega\coloneqq 4/3$.

The potential has become an ordinary quadratic well, so we can use standard
techniques from the quantum harmonic oscillator to solve the Schr\"{o}dinger
equation. Furthermore, since the width of the barrier $r(x-1/2)$ is proportional
to $\epsilon^{1-\alpha}$ and the height is proportional to $\epsilon^{1-\beta}$,
in the region of the barrier, it will overshadow the quadratic potential in the
small $\epsilon$ limit if $(\epsilon^{1-\alpha})^2<\epsilon^{1-\beta}$ which
translates to $1>2\alpha-\beta$. If we restrict ourselves to $\alpha<1/2$ and
$\beta>0$, this is always true, so we can treat the barrier as the dominant
factor in the region where $|x|=O(\epsilon^{1-\alpha})$. Therefore, we can say
that the following is a good approximation for our problem in the large $n$
limit:
\begin{equation}
      \label{eq:quad_app}
      \ppder{\psi}{x} = \epsilon^{-2}\left[V(x)-\epsilon c E\right]\psi(x)
\end{equation}
where
\begin{equation}
      \label{eq:quad_pot}
      V(x)=\begin{cases}
            \epsilon^{1-\beta}&\text{if~}-a<x<a\\
            \omega^2 x^2&\text{otherwise}
      \end{cases},
\end{equation}
where $a\coloneqq \frac{1}{2}\epsilon^{1-\alpha}$. In Eq.~\ref{eq:quad_pot} we
have settled on a form of $r(q)$ that is just a step function. We have focused
on the step function barrier since it makes the differential equation in
Eq.~\ref{eq:quad_app} easy to solve, but we have done numerics that indicate 
other barrier shapes, such as binomial or Gaussian barriers, give similar 
scaling results for $\gmin$.

\section{Asymptotic Expansion}
\label{sec:asymp}
In this section, we focus on the differential equation in
Eq.~\ref{eq:quad_app} and find the spectral gap. Since Eq.~\ref{eq:quad_app}
describes our original $n$ dimensional hypercube problem in the large $n$ limit,
an asymptotic analysis of Eq.~\ref{eq:quad_app} in the small $\epsilon$ limit
will give us the correct asymptotics for the original problem.

Outside of the barrier, the Schr\"{o}dinger equation looks like that of an
ordinary quantum harmonic oscillator problem, but we cannot use the standard
harmonic oscillator solutions since these have already had boundary conditions
imposed, ensuring that the wave-functions go to zero as $x\to\pm\infty$. To get
the solutions for arbitrary boundary conditions, we can compare the harmonic
oscillator equation to the Weber equation \cite{Mathworld}
\begin{equation}
\dder{D_\nu(z)}{z} +\left(\nu+\frac{1}{2}-\frac{1}{4}z^2\right)D_{\nu}(z)=0,
\end{equation}
where $\nu$ is an arbitrary eigenvalue, and $D_{\nu}(z)$ is known as a parabolic
cylinder function. Note that when $\nu$ is a positive integer and $z$ is real,
these functions become the standard Gaussians times Hermite polynomials we
expect from the harmonic oscillator. 

When $\nu$ is not a positive integer and $z$ is real, these functions blow up as
$z\to-\infty$ but go to zero as $z\to\infty$, so we can use these as the
solution to our DE for $x>a$. Furthermore, to get the solution in the $x<-a$
region, we can just employ the symmetry of our problem about $x=0$ to say that
we either have symmetric or anti-symmetric eigenfunctions. Therefore, the
eigen-solutions to our differential equation will have the form

\begin{equation}
\psi(x) = 
\begin{cases}
\pm A_1 D_{\nu_\pm}\left(-\sqrt{\frac{2\omega}{\epsilon}}x\right)
& \text{if~}x<-a\\
A_2 e^{k_\pm x}\pm A_2 e^{-k_\pm x}
&\text{if~} -a<x<a\\
A_1 D_{\nu_\pm}\left(\sqrt{\frac{2\omega}{\epsilon}}x\right)
& \text{if~}x>a
\end{cases},
\end{equation}
where $\nu_\pm\coloneqq \frac{cE_{\pm}}{2\omega}-\frac{1}{2}$ and
$k_{\pm}\coloneqq \sqrt{\epsilon^{-1-\beta}-\epsilon^{-1} cE_{\pm}}$.

By applying continuity in the wave-function and its derivative across the
boundary at $x=\pm a$, we can find a transcendental equation for the energies,
which we denote by $E_{\pm}$ representing the two lowest level energy
states:
\begin{align}
      \label{eq:trans_con}
      k_\pm D_{\nu_\pm}\left(\sqrt{\frac{2\omega}{\epsilon}}a\right)\left(e^{k_\pm a}\mp e^{-k_\pm a}\right)
            &=\sqrt{\frac{2\omega}{\epsilon}} D'_{\nu_\pm}\left(\sqrt{\frac{2\omega}{\epsilon}}a\right)\left(e^{k_\pm a}\pm e^{-k_\pm a}\right).
\end{align}

\begin{figure}
      \includegraphics[width=0.48\textwidth]{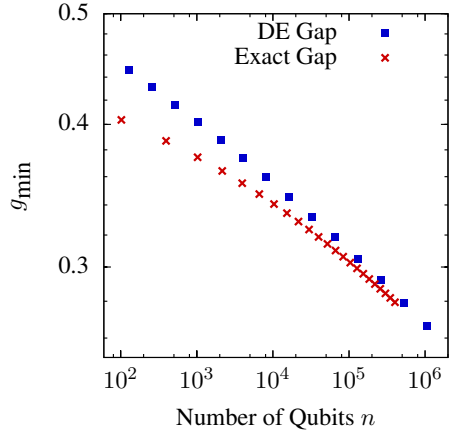}
      \caption{            
                  Comparison between the scaling of the true spectral gap obtained by
                  diagonalization of Eq.~\ref{eq:tridiag} and the gap obtained by solving 
                  for the eigenenergies of the differential equation in Eq.~\ref{eq:quad_app}. 
                  The latter is calculated by numerically solving the transcendental equation in
                  Eq.~\ref{eq:trans_con}. 
     Note that these scalings converge for $n>10^5$, confirming that our derivation of Eq.~\ref{eq:quad_app}
     is indeed valid in the limit of large $n$. 
     This data was obtained for a rectangular barrier with $\alpha=\beta = 0.3$.
      }
      \label{fig:trans-actual}
\end{figure}

This transcendental equation can be solved numerically for the lowest energy levels, and a comparison of this numerical solution to the full spectral gap of the Hamiltonian in Eq.~\ref{eq:tridiag} is shown in Fig.~\ref{fig:trans-actual}.  In the rest of this section, we show that we can do better than numerical solutions to Eq.~\ref{eq:trans_con} by finding an asymptotic solution in the limit of large $n$.

We expect the energies to be close to the unperturbed first excited state energy
of $E_1=3\omega/c$, so we say $E_\pm = (3\omega+\delta_\pm)/c$ and find
$\delta_\pm$ in the limit of small $\epsilon$. In this limit $\nu_\pm =
\frac{\delta_\pm}{2\omega}+1$ and $k_\pm \approx
\epsilon^{-\frac{1}{2}-\frac{\beta}{2}}$.

At this point, we want to calculate $\gmin = |\delta_+-\delta_-|/c$ up to
leading order in $\epsilon$ using these approximations. If we assume that
$2\alpha+\beta<1$, $\alpha<1/2$, and $\alpha+\beta>1/2$ (this corresponds to the
polynomial region in Fig.~\ref{fig:regions}), then the gap becomes
\begin{equation}
      \label{eq:gap_app}
      \gmin = \frac{8(\omega)^{3/2}}{c\sqrt{\pi}} \epsilon^{\alpha+\beta-\frac{1}{2}}\propto n^{\frac{1}{2}-\alpha-\beta}.
\end{equation}
Similarly, if we assume $2\alpha+\beta>1$ and $\alpha<1/2$, then the gap becomes
\begin{equation}
      \label{eq:gap_exp}
      \gmin = \frac{16(\omega)^{3/2}}{c\sqrt{\pi}} \epsilon^{\beta/2}\text{exp}\left(-\epsilon^{\frac{1}{2}-\alpha-\frac{\beta}{2}}\right)\propto n^{-\frac{\beta}{2}}\text{exp}\left(-(n/2)^{\alpha+\frac{\beta}{2}-\frac{1}{2}}\right).
\end{equation}
This result matches the exponentially small gap found by Kong and Crosson 
\cite{Kong} and Jiang et al. \cite{Jiang}.

The dependences on $\epsilon$ in Eqs.~\ref{eq:gap_app} and \ref{eq:gap_exp} are
exactly what we would expect given Eq. \ref{eq:gold_pred}. Notice that we do not
need to assume $\alpha<\beta$ as in Eq.~\ref{eq:gold_pred}, so our result
extends farther than Goldstone's result and covers the entire area bounded by
$0<\alpha<1/2$ and $\alpha+\beta>1/2$.

\begin{figure}
      \includegraphics[width=0.48\textwidth]{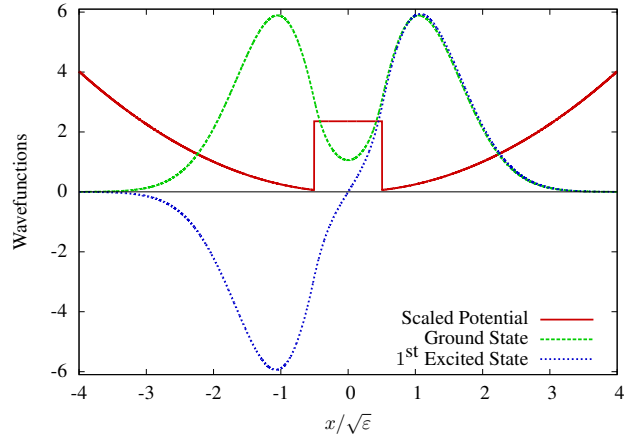}
      \caption{            
                  The ground state and first excited state wavefunctions for the
                  quadratic well approximation of Eq.~\ref{eq:quad_app}. We also
                  display the potential energy from Eq.~\ref{eq:quad_pot}
                  multiplied by a factor of $10/\sqrt{\epsilon}$ so that it is
                  fully visible. Notice that the ground state looks like a
                  Gaussian with the middle dragged downward and the first
                  excited state looks unchanged from the unperturbed quantum
                  harmonic oscillator since the barrier sits in a region where
                  this function was already small. 
      }
      \label{fig:wavefunctions}
\end{figure}

In Fig.~\ref{fig:wavefunctions}, we plot the exact ground state and first
excited state for the quadratic approximation for $\epsilon=1/5000$, a bump
width of $1/70$, and a bump height of $1/300$. These values were chosen to
provide visibility for the bump and its effect on the eigenfunctions. The
potential is also plotted, multiplied by $10/\sqrt{\epsilon}$ so that it is
visible. Here we are using the exact energies, obtained by solving the
transcendental equation, Eq.~\ref{eq:trans_con}, numerically. Notice that the
ground state looks like a Gaussian with its center pulled down whereas the first
excited state looks almost unchanged from its unperturbed state. The first
excited state is unchanged because it was already small in the vicinity of
$x=0$, so the barrier does not alter this state much by making that region more
unfavorable. This is also reflected in our approximation in Eq.~\ref{eq:gap_app}
where the leading order term shown here is due to the ground state rather than
the first excited state.

\section{Conclusion}
\label{sec:conc}
We have taken our original $n$ qubit barrier tunneling problem and, through a
series of large-$n$ approximations, have arrived at an elementary barrier
tunneling problem in one continuous dimension. The resulting approximate
Schr\"{o}dinger equation gives a transcendental equation for the energies which
in the large-$n$ limit gives a spectral gap that is proportional to
$1/n^{\alpha+\beta-\frac{1}{2}}$ for $\alpha+\beta>1/2$ and $2\alpha+\beta<1$
and $n^{-\beta/2} e^{-Cn^{(2\alpha+\beta-1)/2}}$. Our gap scaling result
verifies and provides a solid proof the folklore result by Goldstone
\cite{Goldstone}.

Combined with the work of Reichardt \cite{Reichardt} and Kong and Crosson
\cite{Kong}, our result provides a full picture of the asymptotic behavior of
the spectral gap during barrier tunneling for a symmetric cost function on $n$
qubits. Additionally, our method holds no matter where the barrier is centered
(with suitable redefinitions of $c$ and $\omega$). Our work does focus on a step
function barrier and can therefore be made more general in terms of barrier
shape, but numerics indicate that other barrier shapes give the same scaling for 
$\gmin$.

\subsubsection*{Acknowledgements}
This material is based upon work supported by the National Science Foundation under Grant No. 1314969.


\appendix

\section{Discrete Villain Transformation}
\label{app:villain}

Our goal in this appendix is to fill in the gaps in the derivation of
Eq.~\ref{eq:post_villain}, starting at Eq.~\ref{eq:spin_ham2}. A standard way of
dealing with this problem would be to use the Villain representation
\cite{Villain} which first takes the continuum limit of the eigenvalues of
$\hat{\jmath}_z$ and then defines a conjugate momentum to this continuous
``position'' variable. This technique, as it is commonly implemented, has many
subtleties that our ignored, so in this appendix, we formalize the
assumptions implicit in the Villain representation. Furthermore, we extend
these results and show that certain assumptions about the ground and first
excited states allow us to refine our approximations and create an easier to
understand picture. We also derive all of our results in the discrete case
using linear algebra and only resort to the continuum limit at the end,
elucidating exactly what our assumptions mean.

The original Villain transformation \cite{Villain} says that for
$\hat{\jmath}_z$ and $\hat{\jmath}_\pm \coloneqq \hat{\jmath}_x \pm \i
\hat{\jmath}_y$ we have 
\begin{align}
      \label{eq:villain_cont}
      \hat{\jmath}_z\ket{q} &= q\ket{q}\nonumber\\
      \hat{\jmath}_+\ket{q} &= \e^{-\i p}\sqrt{1+\epsilon-q(q+\epsilon)}\ket{q}\nonumber\\
      \hat{\jmath}_-\ket{q} &= \sqrt{1+\epsilon-q(q+\epsilon)}\e^{\i p}\ket{q}
\end{align}
Here $p$ is the conjugate momentum for $q$, and in $q$-space it can be
represented as $p=-\i\epsilon\pder{}{q}$. Many users \cite{Enz, Boulatov, Boxio}
of this approximation then employ the small $\epsilon$ limit to say that the
square root factors in the $\hat{\jmath}_\pm$ expressions are the same and that
the commutators between $q$ and $p$ are negligible. Using these approximations,
they find that
\begin{equation}
      \label{eq:vill_orig}
      \hat{\jmath}_x = \frac{1}{2}\left(\hat{\jmath}_++\hat{\jmath}_-\right) = \sqrt{1+q^2}\cos{p}.
\end{equation}
It turns out that there are subtleties here, most of them centering around how
big $p$ is. It turns out that this expression is true only to zeroeth order in
$\epsilon$, but because the derivative operator has norm proportional to
$\epsilon^{-1}$, there is a relavant term to second order in $p$. However, this
expression is incorrect at all higher orders of $\epsilon$ and in fact includes
terms linear in $p$ that are not included here, and most misleading, the true
expression contains no terms higher order in $p$ than squared.

 Below, we go through a more formal derivation of $\hat{\jmath}_x$'s
expansion using the underlying matrices and proceed with a description of
how the discrete Villain transformation can be used in the setting of our
problem.

\subsection{Discrete $j$-operators}
We remind the reader that $\epsilon=2/n=1/J$, and we start by examining
$\hat{\jmath}_x$ in the eigenbasis of $\hat{\jmath}_z$ given by $\ket{q}$ where
$q=\epsilon m \in [-1,+1]$ for $m\in\{-J,-J+1,\dots,J\}$. We look at how
$\hat{\jmath}_x$ acts on some general state $\ket{\psi}$ where
\begin{equation}
      \hat{\jmath}_x\ket{\psi} = \sum_{q}\hat{\jmath}_x\psi_q\ket{q}
\end{equation}

We now introduce three new operators such that we can easily represent the 
raising and lowering operators $\hat{\jmath}_\pm$:
\begin{align}
      {\opP} = \sum_{q\in[-1,1-\epsilon]}\ket{q+\epsilon}\bra{q}
      \quad
      \text{and}
      \quad
      {\opM} = \sum_{q\in[-1+\epsilon,1]}\ket{q-\epsilon}\bra{q}
\end{align}
\begin{align*}
      {\opQ} = \sum_{q} q\ket{q}\bra{q}.
\end{align*}
Since ${\opQ}$ is diagonal, functions of it are easy to calculate, and the first
two operators extract just raising and lowering of indices without any
prefactors. Therefore, we can represent our operator as
\begin{align}
      \label{eq:vill_mat1}
      \hat{\jmath}_x  &= \frac{1}{2}(\hat{\jmath}_+ + \hat{\jmath}_-)\\\nonumber
      &= \frac{1}{2}\left(\sqrt{(1-{\opQ})(1+{\opQ}+\epsilon)}{\opP}+\sqrt{(1+{\opQ})(1-{\opQ}+\epsilon)}{\opM}\right).
\end{align}

Our eventual goal is to take a continuum limit of $q$, so we now look
at the matrices that lead to derivatives in this limit that
$\braket{q}{\psi}=\psi_q\to\psi(q)$:
\begin{align}
      \pder{\psi}{q}  &= \lim_{\epsilon\to 0} \frac{\psi(q+\epsilon)-\psi(q-\epsilon)}{2\epsilon}\nonumber\\ 
                      &= \lim_{\epsilon\to 0} \frac{\psi_{q+\epsilon}-\psi_{q-\epsilon}}{2\epsilon}= \lim_{\epsilon\to0} ({\opA}\vec{\psi})_q,\\
      \ppder{\psi}{q} &= \lim_{\epsilon\to 0} \frac{\psi(q+\epsilon)-2\psi(q)+\psi(q-\epsilon)}{\epsilon^2}\nonumber\\
                      &= \lim_{\epsilon\to 0} \frac{\psi_{q+\epsilon}-2\psi_q+\psi_{q-\epsilon}}{\epsilon^2} = \lim_{\epsilon\to 0} ({\opB}\vec{\psi})_q.
\end{align}
Here we have defined two new operators that correspond to the discrete versions
of our first and second derivatives
\begin{align}
      \label{eq:der_ops}
      \epsilon{\opA} = \frac{{\opP}-{\opM}}{2}\quad\text{and}\quad \epsilon^2{\opB} = {\opP}-2{\opI}+{\opM}.
\end{align}
Throughout this appendix, we refer to the relative sizes of certain
operators by refering to their matrix norm. The definition of the matrix norm we
are using is the maximum absolute value of any eigenvalue of the operator.
Therefore, the norm of $\opA$ is proportional to $\epsilon^{-1}$, and the norm
of $\opB$ is proportional to $\epsilon^{-2}$. We come back to revisit this
concept later in the context of our specifical eigenstates.

Notice based on the definitions in Eq.~\ref{eq:der_ops} that we can rewrite
${\opP}$ and ${\opM}$ in terms of ${\opI}$, ${\opA}$, and ${\opB}$:
\begin{align}
      \label{eq:disc_deriv}
      {\opP} = {\opI}+\epsilon{\opA}+\frac{\epsilon^2}{2}{\opB}\quad\text{and}\quad {\opM} = {\opI}-\epsilon{\opA}+\frac{\epsilon^2}{2}{\opB}.
\end{align}
Notice that in the continuous form of the Villain representation in
Eq.~\ref{eq:villain_cont}, these operators, ${\opP}$ and ${\opM}$, correspond to
$e^{\mp \i p}$, but here we see that the operators only correspond to the first
two terms in the Taylor expansion of the exponentials. In addition, the matrix
norms of $\opA$ and $\opB$ further complicate the issue, making it deceptively
appear that the later terms in the series are smaller when in fact every term in
this series is roughly equivalent in size, relative to $\epsilon$.

We can plug the expansions in Eq.~\ref{eq:disc_deriv} back into
Eq.~\ref{eq:vill_mat1} to get the discrete form of the Villain representation.
It should be noted that the following expression is exact and includes no
approximations yet
\begin{align}
      \label{eq:vill_mat2}
            \hat{\jmath}_x = \frac{1}{2}
& \left(\sqrt{(1-{\opQ})(1+{\opQ}+\epsilon)}\left[{\opI}+\epsilon{\opA}+\frac{\epsilon^2}{2}{\opB}\right] +\sqrt{(1+{\opQ})(1-{\opQ}+\epsilon)}\left[{\opI}-\epsilon{\opA}+\frac{\epsilon^2}{2}{\opB}\right]\right)
\end{align}

Next, we start the actual large spin limit by expanding the square root 
prefactors in orders of $\epsilon$
\begin{align}
      \sqrt{(1\mp{\opQ})(1\pm{\opQ}+\epsilon)}&=\sqrt{1-{\opQ}^2}+\epsilon \frac{1\mp{\opQ}}{2\sqrt{1-{\opQ}^2}}+\mathcal{O}(\epsilon^2)
\end{align}
We can use these expansions in Eq.~\ref{eq:vill_mat2} to write our matrix 
equation as
\begin{align}
      \label{eq:vill_mat3}
      \hat{\jmath}_x = \frac{1}{2}&\left(\left[2\sqrt{1-{\opQ}^2}+\frac{\epsilon}{\sqrt{1-{\opQ}^2}}\right] 
            - \epsilon^2\frac{{\opQ}}{\sqrt{1-{\opQ}^2}}{\opA}\right.\nonumber\\
            &+ \left.\epsilon^2\left(\sqrt{1-{\opQ}^2}+\frac{\epsilon}{2\sqrt{1-\opQ^2}}\right){\opB}
            + \mathcal{O}(\epsilon^2)\right).
\end{align}
Here I have kept terms up through terms that are proportional to $\epsilon$,
remembering that the norms of $\opA$ and $\opB$ are proportional to
$\epsilon^{-1}$ and $\epsilon^{-2}$ respectively. The expression in
Eq.~\ref{eq:vill_mat3} can be thought of as a more accurate version of
Eq.~\ref{eq:vill_orig}, and if we were to take only terms that are constant with
repsect to $\epsilon$, we would recover Eq.~\ref{eq:vill_orig}, assuming that we
only take constant terms from that expression as well.

It turns out that some knowledge of our eigenstates can restrict the form of
Eq.~\ref{eq:vill_mat3} even more, so that we can talk about the maximum
eigenvalue of $\opA$ and $\opB$ relevant to the low energy eigenvectors of our
problem, rather than the maximum eigenvalues obtainable for a general problem.
We come back and reexamine the scaling behavior of ${\opA}$ and ${\opB}$.

There are a few key things to note about Eq.~\ref{eq:vill_mat3} in relation to
Eq.~\ref{eq:vill_orig}. First, we only have terms up to the second derivative,
even if we included terms to all orders in $\epsilon$. Second, this form of the
operator makes no assumptions about the specific form of the Hamiltonian or its
energy states. If, as we do in the next section, we make assumptions about our
energy states and problem, we can further this approximation and find an even
simpler form.

\subsection{Hamiltonian and Eigenstate-based Approximations}

Our next step involves putting this into our full equation, where now, the 
Hamiltonian is given by
\begin{equation}
      \epsilon\hat{H} = -(1-s)\hat{\jmath}_x+s\hat{\jmath}_z+s r(\hat{\jmath}_z)+\Delta
\end{equation}

For this, $\hat{\jmath}_z\to{\opQ}$, and we can use our form for
$\hat{\jmath}_x$ in Eq.~\ref{eq:vill_mat3}, interpreting $\vec{\psi}$ as an
eigenstate with eigenenergy $E$, so that our Schr\"{o}dinger equation becomes
\begin{align}
      \label{eq:vill_schro1}
      \epsilon E \ket{\psi} &= \left(s{\opQ}+sr({\opQ})+\Delta-(1-s)\left[\sqrt{1-{\opQ}^2}+\frac{\epsilon}{2\sqrt{1-{\opQ}^2}}\right]\right.\nonumber\\
                            &\left.+(1-s)\frac{\epsilon^2}{2}\frac{{\opQ}}{\sqrt{1-{\opQ}^2}}{\opA}\right.\nonumber\\
                            &\left.-(1-s)\frac{\epsilon^2}{2}\left(\sqrt{1-{\opQ}^2}+\frac{\epsilon}{2\sqrt{1-\opQ^2}}\right){\opB}+ \mathcal{O}(\epsilon^2)\right)\ket{\psi}.
\end{align}
A version of this equation is shown in Eq.~\ref{eq:schro_vill} where we have
taken the continuum limit, treating $q$ as a continuous variable so that
${\opQ}\to q$, $\vec{\psi}\to \psi(q)$, ${\opA}\to \pder{\psi}{q}$, and
${\opB}\to \ppder{\psi}{q}$.

Next, we define the operator ${\opx} = {\opQ}+\frac{1}{2}$, and we 
call the diagonal entries of this operator $x = q+\frac{1}{2}$. Our reasoning in
defining this is to get a variable that is small in the vicinity of
$q=-\frac{1}{2}$ where our tunneling event is going to happen. We also go
ahead and put ourselves at the critical $s^* = \frac{1}{2}(\sqrt{3}-1)$.

Now, we are going to look more closely at ${\opx}$ and its relationship to
$\ket{\psi}$. The low energy eigenvectors, $\ket{\psi}$, will essentially be
zero for most of their entries except right around the location of the primary
bump in the distribution. The reasoning behind this comes from the fact that for
the low lying energy states, their energy is lower than the potential energy
function for the entire range of $x$, except in an extremely narrow range around
the barrier, leading to exponential suppresion of the wavefunctions outside this
region. For the no barrier case, the ground state and first excited state both
have width $\mathcal{O}(\sqrt{\epsilon})$ and are centered around $x=0$ with
exponential supression away from $x=0$.

Since the widths of the ground state and first excited state
($\mathcal{O}(\sqrt{\epsilon})$) are larger than the width of the barrier
($\mathcal{O}(\epsilon^{1-\alpha})$), the range of $x$ over which the components
of $\ket{\psi}$ are non-zero is $\mathcal{O}(\sqrt{\epsilon})$.
Therefore, focusing on the diagonal terms in the Schr\"{o}dinger equation that
do not include ${\opA}$ or ${\opB}$, we can expand these to order $\epsilon$ by
treating $||{\opx}\ket{\psi}||\in \mathcal{O}(\sqrt{\epsilon})$ since for the
entries of $\ket{\psi}$ that matter, the typical $x$ values will be of order
$\sqrt{\epsilon}$. We also use the arbitrary constant $\Delta$ to cancel
out the constant terms in the expansion, physically ensuring that the bottom of
our potential well is at zero energy:
\begin{align}
      \left(s^*{\opQ}+\Delta-(1-s^*)\left[\sqrt{1-{\opQ}^2} +\frac{\epsilon}{2\sqrt{1-{\opQ}^2}}\right]\right)\ket{\psi}& 
      = \left(\frac{2}{3}(\sqrt{3}-1){\opx}^2\right)\ket{\psi}+\mathcal{O}(\epsilon^{3/2}),
\end{align}
where $\Delta = -\frac{(\sqrt{3}-1)}{24}(24+12\epsilon)$. Next, we focus on
the derivative terms. We would expect the size of the derivative to be governed
by the inverse of the length scale over which the eigenvector components change.
In the unperturbed case, we would expect the eigenvector (which is a binomial
distribution) components to change on a length scale of $\sqrt{\epsilon}$ which
would mean that ${\opA}$ scales like $1/\sqrt{\epsilon}$ and ${\opB}$ scales
like $1/\epsilon$. Note that these would then correspond to the norms
$||\opA\ket{\psi}||$ and $||\opB\ket{\psi}||$ not $||\opA||$ and $||\opB||$
which as we discussed in the last section can be much larger. This scaling
behavior requires our restriction to the low-lying energy states.

In the perturbed case, we expect the shortest length scale in the problem to be
governed by the exponential decay inside the barrier. In the prototypical
barrier tunneling problem of plane waves tunneling through a square barrier, the
Schr\"{o}dinger equation inside the barrier, which will have extent
$-\xi<y<\xi$, will be of the form
\begin{equation}
      \label{eq:reg_barrier}
      \dder{\varphi}{y} = \frac{2}{\hbar^2}(V_0-\mathcal{E})\varphi(y) = k^2\varphi(y),
\end{equation}
where in our problem $\hbar\to\epsilon$, the width of the barrier $\xi$ is
proportional to $\epsilon^{1-\alpha}$, $V_0$ is the height of the barrier which
for us is proportional to $\epsilon^{1-\beta}$, and $\mathcal{E}$ is the energy
which in our small $\epsilon$ limit should be much smaller than $V_0$. If we
compare this to our expression in Eq.~\ref{eq:vill_schro1}, we see that at the
very least there is still a factor of $\epsilon^2$ in the ratio between the
potential barrier $r(q)$ and the second derivative term $\opB$. This is a rough
comparison, but we can use it to inform what the exponential decay inside the
barrier looks like.

Specifically, Eq.~\ref{eq:reg_barrier} is solved by $e^{\pm k y}$, where in our
case $k$ is proportional to $\epsilon^{-\frac{1}{2}-\frac{\beta}{2}}$, assuming
that $V_0\gg \mathcal{E}$, which is a good assumption in our problem. Therefore,
the length scale over which the wavefunction changes inside the barrier is
proportional to $\epsilon^{\frac{1}{2}+\frac{\beta}{2}}$. This means that we can
claim our derivative, and therefore $\opA$, scales like
$\epsilon^{-\frac{1}{2}-\frac{\beta}{2}}$, and similarly $\opB$ scales like
$\epsilon^{-1-\beta}$. Alternatively, the derivatives are proportional to $k$
and $k^2$ for $\opA$ and $\opB$ respectively.

One other thing to note is that if we are looking at a derivative that scales
like $\epsilon^{\frac{1}{2}+\frac{\beta}{2}}$, it will only be doing this in the
region close to the edge of the barrier, which means that
$x\in\mathcal{O}(\epsilon^{1-\alpha})$ when we care about derivatives that are
this large. Thus in keeping track of the order, we need to remember that
higher order terms in $x$ will be even more exacerbated in this region of the
barrier.

At the moment, we focus on just the lowest order terms that include $\opA$
and $\opB$. Using these approximations, we get for the operator $\opA$:
\begin{align}
      (1-s^*)\epsilon^2\frac{{\opQ}}{2\sqrt{1-{\opQ}^2}}{\opA} & =  -\frac{1}{2} \left(\sqrt{3}-1\right) \epsilon ^2{\opA} + \mathcal{O}(\epsilon^{\frac{5}{2}-\alpha-\frac{\beta}{2}}) = \mathcal{O}(\epsilon^{\frac{3}{2}-\frac{\beta}{2}}), 
\end{align}
while for $\opB$ we have
\begin{align}
      -(1-s^*)\frac{\epsilon^2}{2}\left(\sqrt{1-{\opQ}^2}+\frac{\epsilon}{2\sqrt{1-\opQ^2}}\right){\opB} &=
      -\frac{3}{8} \left(\sqrt{3}-1\right) \epsilon ^2{\opB} + \mathcal{O}(\epsilon^{2-\alpha-\beta}) = \mathcal{O}(\epsilon^{1-\beta})
\end{align}

Our condition for whether a term is small or not depends on the energy term.
Our unperturbed energies are constant with $\epsilon$, but notice that $E$ is
multiplied by $\epsilon$ in the Schr\"{o}dinger equation. Thus, we expect this
energy term to be proportional to $\epsilon$ with some polynomially or
exponentially small corrections. Therefore, if a term is higher order than
linear in $\epsilon$, we discard it since it is smaller than the energy
term which is what we care about.

We assume $\beta<1$, in which case, the $\opA$ terms are all small. For the
$\opB$ terms, we see that we need $1<2-\alpha-\beta$ in order for the next
highest term to contribute, so we need to restrict ourselves to
$\alpha+\beta<1$. The only remaining thing to look at is how much the
$r({\opQ})$ term will contribute. The height of the barrier scales like
$\epsilon^{1-\beta}$ in this setup, so as long as $\beta>0$, we are able to
keep the barrier as well. Going through all of this work, our Schr\"{o}dinger
equation becomes
\begin{align}
      \epsilon E \vec{\psi} & = \left(s^*r\left({\opx}-\frac{1}{2}\right)+\frac{2}{3}(\sqrt{3}-1){\opx}^2 
      - \frac{3}{8} \left(\sqrt{3}-1\right) \epsilon ^2{\opB} + \mathcal{O}\left(\max\{\epsilon^{2-\alpha-\beta},\epsilon^{\frac{3}{2}-\frac{\beta}{2}}\}\right)\right)\vec{\psi}
\end{align}

Now, we are finally in a good position to do the continuum limit, taking $x$ to
a continuous variable in the small $\epsilon$ limit, and taking ${\opB} \to
\ppder{}{x}$, ${\opx}\to x$, and $\ket{\psi} \to \psi(x)$. Doing this gives the
differential equation

\begin{align}
      \epsilon E \psi(x) & = \left(s^*r\left(x-\frac{1}{2}\right)+\frac{2}{3}(\sqrt{3}-1)x^2 - \frac{3}{8} \left(\sqrt{3}-1\right) \epsilon ^2\ppder{}{x} + \mathcal{O}\left(\max\{\epsilon^{2-\alpha-\beta},\epsilon^{\frac{3}{2}-\frac{\beta}{2}}\}\right)\right)\psi(x)
\end{align}

This can be solved using the parabolic cylinder equations we use in the main
portion of the paper.

\end{document}